\documentclass[structabstract]{aa}

\usepackage{graphicx}
\usepackage{txfonts}
\usepackage[caption=false]{subfig}
\usepackage{natbib}
\bibpunct{(}{)}{;}{a}{}{,}

\begin{document}

\title {Using Galactic Cepheids to verify Gaia parallaxes}
\author{{F. Windmark}\inst{\ref{inst1},\ref{inst2}}\and {L. Lindegren}\inst{\ref{inst2}} \and {D. Hobbs}\inst{\ref{inst2}}}

\institute{Max-Planck-Institut f{\"u}r Astronomie, K{\"o}nigstuhl 17, 69117 Heidelberg, Germany \label{inst1} \\ \email{windmark@mpia.de} \and Lund Observatory, Box 43, 221 00 Lund, Sweden\label{inst2}}

\date{Received 20 Mars 2011 / Accepted 8 April 2011}

\abstract{The Gaia satellite will measure highly accurate absolute parallaxes of hundreds of millions of stars by comparing the parallactic displacements in the two fields of view of the optical instrument. The requirements on the stability of the `basic angle' between the two fields are correspondingly strict, and possible variations (on the microarcsec level) are therefore monitored by an on-board metrology system. Nevertheless, since even very small periodic variations of the basic angle might cause a global offset of the measured parallaxes, it is important to find independent verification methods.}
{We investigate the potential use of Galactic Cepheids as standard candles for verifying the Gaia parallax zero point.}
{We simulate the complete population of Galactic Cepheids and their observations by Gaia. Using the simulated data, simultaneous fits are made of the parameters of the period--luminosity relation and a global parallax zero point.}
{The total number of Galactic Cepheids is estimated at about 20\,000, of which nearly half could be observed by Gaia. In the most favourable circumstances, including negligible intrinsic scatter and extinction errors, the determined parallax zero point has an uncertainty of 0.2~microarcsec. With more realistic assumptions the uncertainty is several times larger, and the result is very sensitive to errors in the applied extinction corrections.}
{The use of Galactic Cepheids alone will not be sufficient to determine a possible parallax zero-point error to the full potential systematic accuracy of Gaia. The global verification of Gaia parallaxes will most likely depend on a combination of many different methods, including this one.}

\maketitle

\section{Introduction}

The Gaia satellite, due for launch in 2013, will measure the trigonometric parallaxes of roughly a billion objects in the Galaxy and beyond with accuracies reaching 10~$\mu$as
\citep[microarcsec;][]{2010IAUS..261..296L}. This huge improvement over its predecessor Hipparcos, in terms of accuracy, limiting magnitude and number of objects, will revolutionize many areas of stellar and Galactic astrophysics \citep{2001A&A...369..339P}. Moreover, it will allow entirely new kinds of investigations that depend on the statistical combination of very large data sets. One such example is the determination of the distance to the Large Magellanic Cloud (LMC). The LMC distance is fundamental for the extragalactic distance scale, and current estimates put it at 50~kpc (or 20~$\mu$as parallax) with a relative uncertainty of 5\% \citep{2001ApJ...553...47F,2008AJ....135..112S}. Gaia should be able to observe some $N\simeq 10^7$ stars brighter than 20th magnitude in the LMC, with a standard error in the individual parallaxes of about 200~$\mu$as or better. Potentially, therefore, the mean LMC parallax as estimated from Gaia data could have an accuracy of $200 N^{-0.5} \simeq 0.06~\mu$as, equivalent to a relative error in distance of 0.3\%, or 0.006 in distance modulus. 
Needless to say, such a result will be extremely interesting. 
However, to achieve the $N^{-0.5}$ improvement for large $N$ requires (1) that the individual parallax errors are effectively uncorrelated, which may be the case \citep{2010IAUS..261..320H}, and (2) that there is no significant global zero-point error (bias) in the measured parallaxes. This and many similar examples show that even a bias $<0.1~\mu$as must be considered significant in comparison with the potential capabilities of the mission.   

Achieving the desired parallax accuracy requires an exceedingly stable optical instrument in the Gaia satellite. Even extremely small periodic variations in the so-called basic angle between the two fields of view could lead to an undesirable global offset of the measured parallaxes \citep{2011AdSpR..47..356M}. 
One well-known cause of such variations is the variable heating from solar radiation as the satellite rotates. Although the satellite has been carefully designed to minimize these effects, the resulting variations in the basic angle cannot be completely eliminated. With the use of an on-board laser interferometer they will however be continuously measured and taken into account in the instrument calibration model. However, it is obviously of great importance to be able to verify the resulting parallax zero point by independent, astrophysical means.

A large number of methods are in principle available for astrophysical verification of the Gaia parallaxes. We may distinguish three main classes of methods: (1) A priori knowledge of the parallax, using for example quasars that are so distant that their parallaxes for the present purpose can be considered to be zero. (2) Distances determined by geometric principles not relying on trigonometric parallax. Several of these methods depend on a combination of doppler velocity, time and angular measurements, for example orbital parallaxes for spectroscopic binaries with an astrometric orbit \citep[e.g.,][]{1997ApJ...485..167T}, expansion parallaxes for supernova remnants and planetary nebulae \citep[e.g.,][]{1973PASP...85..579T,2002AJ....123.2676L}, a geometric variant of the Baade--Wesselink method for pulsating stars \citep[e.g.,][]{2002ApJ...573..330L}, kinematic distances to globular clusters \citep[e.g.,][]{2006A&A...445..513V,2006ApJS..166..249M}, and rotational parallaxes for external galaxies \citep{2007MNRAS.378.1385O}. (3) Distance ratio methods: the ratio of the distances to two or more objects equals the inverse ratio of their parallaxes. This equality is violated in the presence of a parallax zero point error, which can therefore be derived, e.g., from photometric distance ratios established by means of standard candles, provided that extinction effects can be mastered.

Quasars (belonging to class 1 above) are among the more promising candidates. It is estimated that Gaia will observe around $500\,000$ quasars with individual parallax uncertainties around 250--300 $\mu$as due to their faintness in the Gaia $G$ band \citep{2008IAUS..248..217L}. This would lead to an uncertainty in the mean zero-point around $0.4$ $\mu$as, with a possible additional bias from foreground stars contaminating the sample. 

Distance determinations using various geometric principles (class 2 above) will be very important for checking the consistency and reliability of Gaia parallaxes. However, they are not entirely free of model assumptions, and are therefore in many respects problematic as a means of verifying the Gaia data. Moreover, it is doubtful if the number of objects available and the achievable accuracies are high enough for the present purpose.

In this paper we concentrate on the distance ratio method, based on Classical Cepheids as one of the most reliable standard candles. In particular, we focus on the Galactic Cepheids, which are fewer than the quasars and extragalactic Cepheids but with individually more accurate parallaxes. In this method, to avoid a circular argument, we make a simultaneous calibration of the period--luminosity (P-L) relation and the parallax zero-point. Since Gaia will observe at least ten times the $\sim\,$800 currently known Galactic Cepheids, we need to create a synthetic population of Galactic Cepheids with the appropriate properties before we can simulate the Gaia observations and make a statistical investigation of the expected errors.
Since it is difficult to quantify how much of the final errors will depend on various modelling errors (say, from a possible non-linearity of the P-L relation), our strategy is to consider first a best-case (i.e., optimistic) scenario, where all such effects are negligible, and then investigate how sensitive the results are to the various assumptions.

The method of absolute parallax measurements with Gaia is closely related to the physical origin of a possible parallax bias, and these aspects of the mission are therefore briefly explained in Sect.~\ref{sec:gaia}. Our modelling of the Galactic Cepheids and their observations by Gaia are described in Sect.~\ref{sec:modelling}. The results of the simulation and parameter fitting experiments are discussed in Sect.~\ref{sec:analysis}, followed by the conclusions in Sect.~\ref{sec:conclusions}.

\begin{figure}
\centering
\includegraphics[clip=true, width = 7cm]{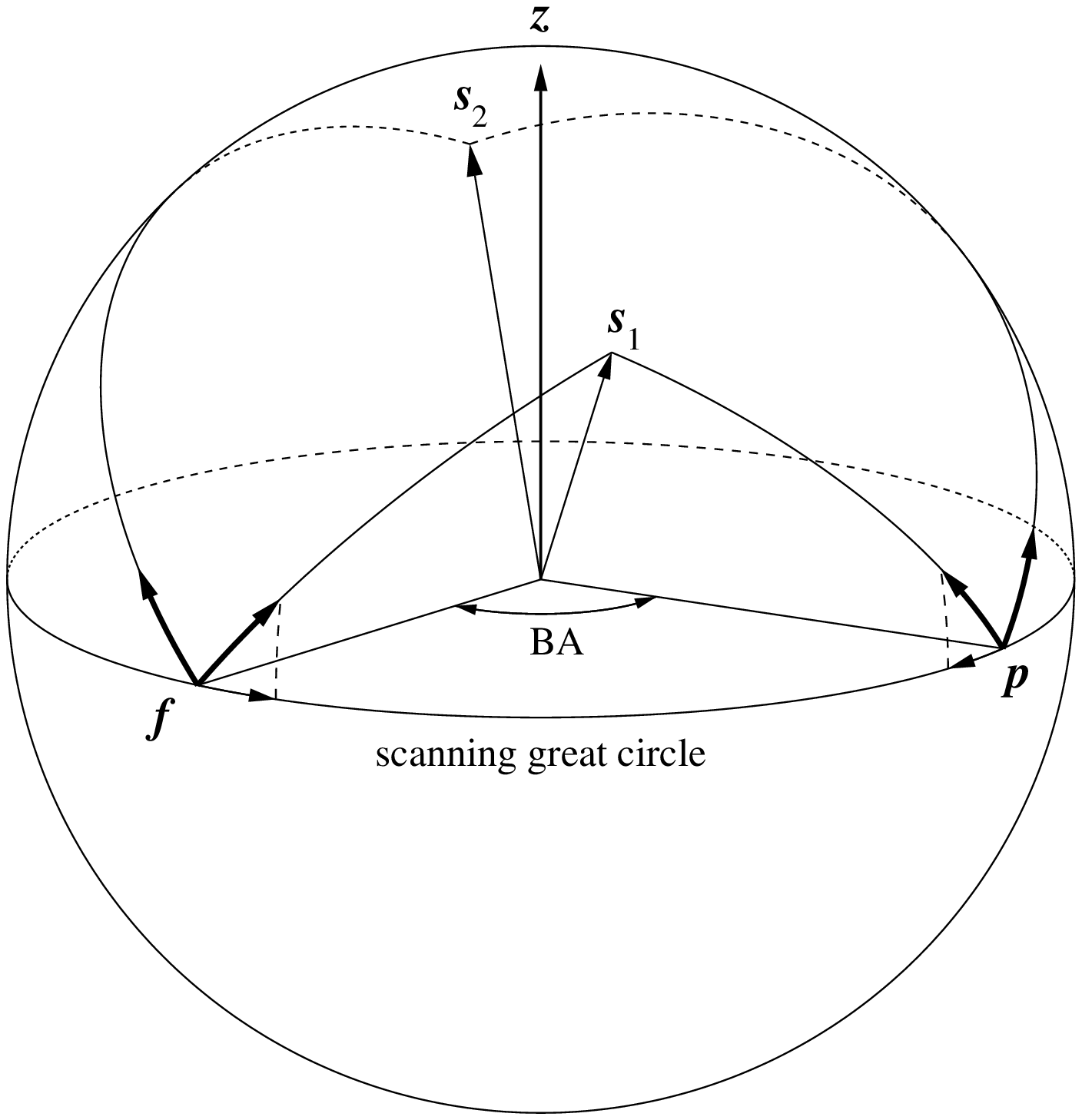}
\caption{ \label{fig:ba} Illustrating the effect of parallax on stellar images
in the preceding ($\vec{p}$) and following ($\vec{f}$) field of view of Gaia,
when the Sun is in two different positions ($\vec{s}_1$ and $\vec{s}_2$)
relative to the stars. $\vec{z}$ is the spin axis of Gaia, and the basic angle is
marked between the two fields of view. The displacement due 
to parallax is always directed towards the Sun, as indicated by the thick 
arrows at $\vec{p}$ and $\vec{f}$, and is measured by Gaia as projected along the scanning great circle. See Sect.~\ref{sec:gaia} for further explanation.}
\end{figure}

\section{Cause of a possible parallax bias in Gaia} 
\label{sec:gaia}

In contrast to ground-based parallaxes, which are always measured relative to background objects, astrometric satellites such as Hipparcos and Gaia in principle allow to determine \emph{absolute} parallaxes thanks to the large difference in parallax factor between the two widely separated fields of view \citep{2005ESASP.576...29L,2005ASPC..338...25L}. However, this capability depends critically on the short-term ($\sim\,$few hours) stability of the so-called basic angle between the two fields of view \citep{2005A&A...439..805V,2011AdSpR..47..356M}. 
The optical instrument of Gaia is designed to be stable on these time scales to within a few $\mu$as, and an on-board interferometric metrology system, the Basic Angle Monitor (BAM), will moreover measure short-term variations of the basic angle with a precision 
$<\!1~\mu$as every few minutes \citep{2008IAUS..248..217L}. 
The technical design thus guarantees that measurement biases related to basic-angle variations are practically negligible in comparison to the random measurement errors ($\gtrsim 30~\mu$as per field-of-view crossing). Nevertheless, when averaging over many stars we should be concerned about systematic effects that are much smaller than the random errors.

The geometry of the observations with respect to the Sun is important
both for the determination of parallax and for possible thermal perturbations 
of the instrument. Indeed, as discussed by  \cite{2005A&A...439..805V}, a certain
systematic variation of the basic angle, depending on the satellite
spin phase relative to the Sun, has almost the same effect on the
measurements as a global zero-point shift of the parallaxes.
Thus, if such a variation exists in the real Gaia instrument, and is not recognized by the on-board metrology,
then the result will be a global bias of the derived parallaxes. The cause of this
shift can be understood by means of Fig.~\ref{fig:ba}. Stellar parallax causes an
apparent shift of the star along a great circle towards the Sun. If all stars have
a constant positive parallax, the apparent shifts will be as indicated by the 
short arrows in the figure, depending on the position of the Sun relative
to the stars. Thanks to its one-dimensional measurement principle, Gaia is 
only sensitive to the relative shift along the scanning great circle 
through $\vec{p}$ and $\vec{f}$. Thus, with the Sun at $\vec{s}_1$ in the 
diagram, i.e., closest to the fields of view, the stellar parallax shifts on the 
detectors will be indistinguishable from a slight enlargement of the basic angle. 
With the Sun at $\vec{s}_2$, furthest away from the fields, the stellar shifts 
will be indistinguishable from a slight reduction of the basic angle. Thus,
a temporal variation of the basic angle, caused for example by 
the solar heating, would mimic a global shift of the parallaxes.

According to the technical specifications of Gaia, the expected parallax bias
from this effect is at most a few $\mu$as, and most of it will be removed
in the data processing by means of the BAM metrology data. Indeed, the 
BAM specifications are such that, theoretically, the remaining parallax bias 
should be much less than $0.1~\mu$as. Nevertheless,
due to the subtle nature of the effect, it is important to verify the parallax zero point 
by some independent method. Since the relevant instrumental effects are almost completely correlated
with a parallax shift, an independent verification must be based on astrophysical considerations.

\section{Modelling the Cepheids observed by Gaia}
\label{sec:modelling}

We create a synthetic population of Galactic Cepheids with each object assigned a number of properties generated from the relevant probability distribution models. In order to study the Cepheid P-L relation, each object is given a period ($P$) and an absolute visual magnitude ($M_V$), as well as a $V-I$ colour as required for the transformation to the Gaia wide-band $G$ magnitude. Finally, each Cepheid is given a position in the Galaxy from a given distribution model. We base our modelling on the observed distribution of the 455 Cepheids in the \citet{2000A&AS..143..211B} catalogue.

\begin{figure}
\resizebox{\hsize}{!}{ \includegraphics[angle=270]{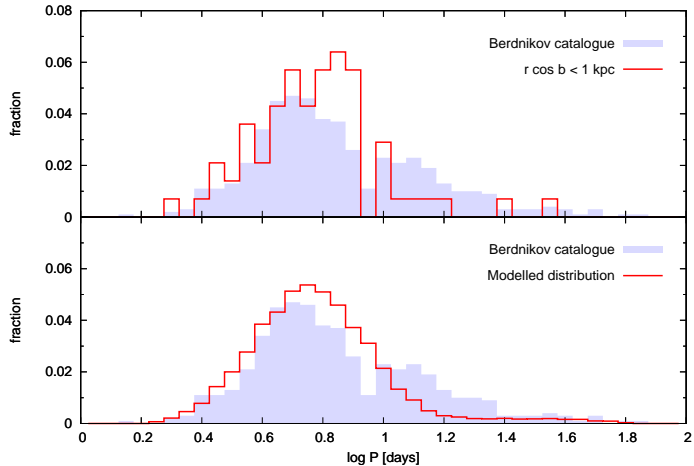} }
\caption{The upper panel shows the period distribution for the Cepheids in the full Berdnikov catalogue (455 Cepheids) as well as the volume-complete sample within the assumed completeness limit of $r \cos b < 1$~kpc (71 Cepheids). Note the larger fraction of long-period Cepheids in the full Berdnikov catalogue due to selection effects. The line histogram in lower panel shows our modelled period distribution consisting of two overlapping Gaussians, which ideally should represent the volume-complete sample in the upper panel. In the model distribution, 96\% of the population has a normal distribution of $\log P$ with mean value 0.75 and standard deviation 0.18, and 4\% has a normal distribution with mean value 1.48 and standard deviation 0.20.} 

\label{fig:perioddist}
\end{figure}

\subsection{Cepheid properties}
\label{sec:properties}

It is now over 100 years since the discovery of the P-L relation \citep{1908AnHar..60...87L, 1912HarCi.173....1L}. It has since played an important part in the extragalactic distance ladder, and its calibration is still of great interest today \citep{2004AA...424...43S, 2007AA...476...73F, 2009ApJ...693..691N}. Because of the difficulties involved with observations in the Galactic plane and the larger number of Cepheids known in the Magellanic Clouds, the majority of the calibrations have used Cepheids in the LMC and SMC at least to determine the slope of the P-L relation. The calibrations show a discrepancy between the P-L relations in the different galaxies, which implies that the luminosities also depend on some other property than the pulsation period. With measured metallicities becoming available for an increasing number of Cepheids, recent studies have shown the metallicity to be a likely candidate, although the results are far from conclusive \citep{2008AA...488...25G, 2008A&A...488..731R}. In this work, we generally assume all Cepheids to follow the same standard P-L relation, disregarding a possible metallicity effect. This gives the best-case scenario for the use of the Galactic Cepheids for the parallax verification. In Sect.~\ref{sec:numres} we briefly consider how a metallicity-dependent P-L relation would affect the results. 

Assuming a period distribution model (see below), the period of each Cepheid is first randomly generated and the true visual magnitude then follows from the assumed P-L relation. As the P-L relation will be calibrated simultaneously with the parallax zero-point, the precise P-L relation adopted for the simulations will affect the end results only minimally. We use the relation from \citet{2004AA...424...43S},
\begin{equation}
	M_V = -3.087 \log P - 0.914 \, ,
\end{equation}
where $P$ is given in days. The period--colour relation by \citet{2003A&A...404..423T} is then used in the same way to generate the intrinsic $V-I$ colour,
\begin{equation}
	(V-I)_0 = 0.256 \log P + 0.497 \, .
\end{equation}
In this process we have neglected the intrinsic dispersion of the P-L relation due to the finite width of the instability strip in the underlying period--luminosity--colour relation
\citep{1991PASP..103..933M}. From LMC data \citep{1999AcA....49..201U} the dispersion 
is found to be about 0.16~mag in $M_V$ and 0.11~mag in $M_I$. The Gaia $G$ magnitude being intermediate between $V$ and $I$ for typical Cepheid colours (cf.\ Eq.~\ref{eq:G}),
the dispersion in $M_G$ is presumably intermediate as well. When using the reddening-free Wesenheit index, \citet{1999AcA....49..201U} found a  considerably smaller dispersion of 0.076~mag. 
For our best-case scenario we ignore the dispersion in the P-L relation, assuming (optimistically) that it can be accounted for by appropriate modelling of the full
period--luminosity--colour--(metallicity)--(other factors) relation for the 
nearby Cepheids, using distances from Gaia.
The effects of an intrinsic dispersion are similar to those of an uncertainty
in the correction for extinction, which we do however investigate (Sect.~\ref{sec:fitting}).

For the modelling of the Cepheid periods, we use the \citet{2000A&AS..143..211B} catalogue to obtain a likely period distribution. Because the period is related to the luminosity, and in order to avoid observational biases, we base our model on the 71 Cepheids within the completeness limit discussed in Sect.~\ref{subsec:cephdist}. In the upper panel of Fig.~\ref{fig:perioddist} we plot the normalized distribution of periods both for the full catalogue of 455 Cepheids (shaded histogram) and for the volume-complete sample of 71 Cepheids (line histogram). We note that the full catalogue contains a larger fraction of long-period (i.e., high-luminosity) Cepheids, as can be expected for a sample that is at least partially limited in apparent magnitude. The full catalogue suggests a bimodal distribution, less evident in the volume-complete sample. In order to reproduce both the strong short-period peak and the long-period tail of the distribution, we fit two overlapping Gaussian functions to the volume-complete sample. The resulting model period distribution
is shown as the line histogram in the lower panel of Fig.~\ref{fig:perioddist}. Although the fit to the volume-complete distribution is far from perfect, the model reproduces the gross distribution reasonably well with a simple continuous function.

\begin{figure}
\resizebox{\hsize}{!}{\includegraphics[angle=270]{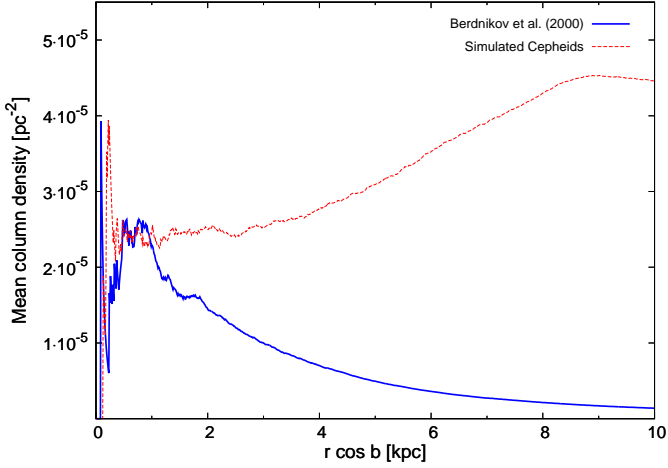}}
\caption{Mean column density of Cepheids within a projected distance $r \cos b$ from the Sun. The Berdnikov Cepheids are given in the solid (blue) curve, and the dashed (red) curve represents simulated data using the radial distribution model. Note the plateau of constant density between 0.5 and 1~kpc indicating that the Berdnikov catalogue is complete within 1~kpc from the Sun.}
\label{fig:coldens}
\end{figure}

\subsection{Spatial distribution of Galactic Cepheids}
\label{subsec:cephdist}

We assume the Cepheids to be distributed axisymmetrically around the Galactic Center, meaning that the number density is only a function of the galactocentric (cylindrical) radius $R$ and the vertical distance $z$ to the Galactic plane:
\begin{equation}
	N(R,z) = \Sigma(R) \times f(R,z) \, ,
\end{equation}
where $\Sigma(R)$ is the surface density at distance $R$ from the axis, and $f(R,z)$ is the density distribution perpendicular to the disk. Since the \citet{2000A&AS..143..211B} catalogue is only complete within $\sim\,$1~kpc around the Sun, it is necessary to estimate the radial distribution by some other means. Since classical Cepheids are young and massive stars, we can expect them to follow roughly the same distribution as other young and bright stars. \citet{1997ApJ...476..144M} and \citet{1997ApJ...476..166W} found the radial distribution of OB associations to be best described by an exponential function $\Sigma(R) \propto \exp(R/R_0)$ with $R_0=3.5$~kpc, and we choose this also for the Galactic Cepheids. From the galactocentric radius, the position is then generated from $x=R\sin\theta$ and $y=R\cos\theta$, where $\theta$ is randomly picked between $0$ and $2\pi$.

The vertical distribution of Cepheids in the Berdnikov catalogue is found to be well fitted by a hyperbolic secant law \citep{1988A&A...192..117V},
\begin{equation}
	f(R,z) = \frac{1}{\pi z_0}~{\rm sech} \Big( \frac{z}{z_0} \Big)
	\label{eq:vertdist}
\end{equation}
where $z_0(R)$ is the radius-dependent scale height. From the Berdnikov catalogue we obtain for the solar neighbourhood $z_0(R = 8.0~\mbox{kpc}) = 75$~pc. \citet{2005AJ....130..659A} found a scale height $z_0 \propto \exp(R/12.5~\mbox{kpc})$ for the distribution of Galactic HI and H$_2$, and we therefore adopt $z_0(R) = (40~\mbox{pc})\times \exp(R/12.5~\mbox{kpc})$ for the Galactic Cepheids.

The surface density of Cepheids at the Sun's distance from the Galactic Centre, $\Sigma(R=8~\mbox{kpc})$, can also be estimated from the \citet{2000A&AS..143..211B} catalogue, if we assume that the Solar neighbourhood is representative and that all Cepheids at high galactic latitudes have been included in the catalogue. In Fig.~\ref{fig:coldens} we plot the mean surface density of the Berdnikov Cepheids within a projected distance $r \cos b$ from the Sun ($r$ is the distance from the Sun and $b$ the Galactic latitude). We note a plateau of roughly constant column density $\simeq 2.5\times 10^{-5}$~pc$^{-2}$ between 0.5 and 1~kpc, before it falls off at larger distances. The plateau is believed to be real, and arises because of the relatively large radial scale length of the Cepheid number density. The fall-off occurs at the distance where the sample is no longer complete. We therefore conclude that the Berdnikov catalogue is complete to a projected distance of about 1~kpc.

To estimate the total number of Galactic Cepheids, we use the radial distribution described above and keep on generating Cepheids until the surface density of the generated sample agrees with the surface density of the Berdnikov sample within the completeness limit. The discrepancy between the dashed (model) and solid (observed) curves in Fig.~\ref{fig:coldens} represents all the Galactic Cepheids that have yet to be detected. This extrapolation results in a total of about $20\,000$ Galactic Cepheids. Changing the radial scale length to $R_0 = 2.5$~kpc, as presented by \citet{1998gaas.book.....B} for the Galactic thin disk, only changes the total number of Cepheids by 10\% up to $22\,000$. These numbers are slightly larger than the $15\,000$ estimated by \citet{2009MNRAS.398..263M}, who did not take the radial gradient into account. Our numbers are in reasonable agreement with expectations from models of star formation and stellar evolution. E.g., assuming that all stars with masses above $5~M_{\sun}$ become Cepheids with a mean lifetime of 2~Myr, a total star formation rate of $3~M_{\sun}~\mbox{yr}^{-1}$, and the \citet{1955ApJ...121..161S} IMF, we expect $\sim{}20\,000$ Cepheids in the Galaxy.

\begin{figure*}
	\centering
	\includegraphics[angle=270, width=13cm]{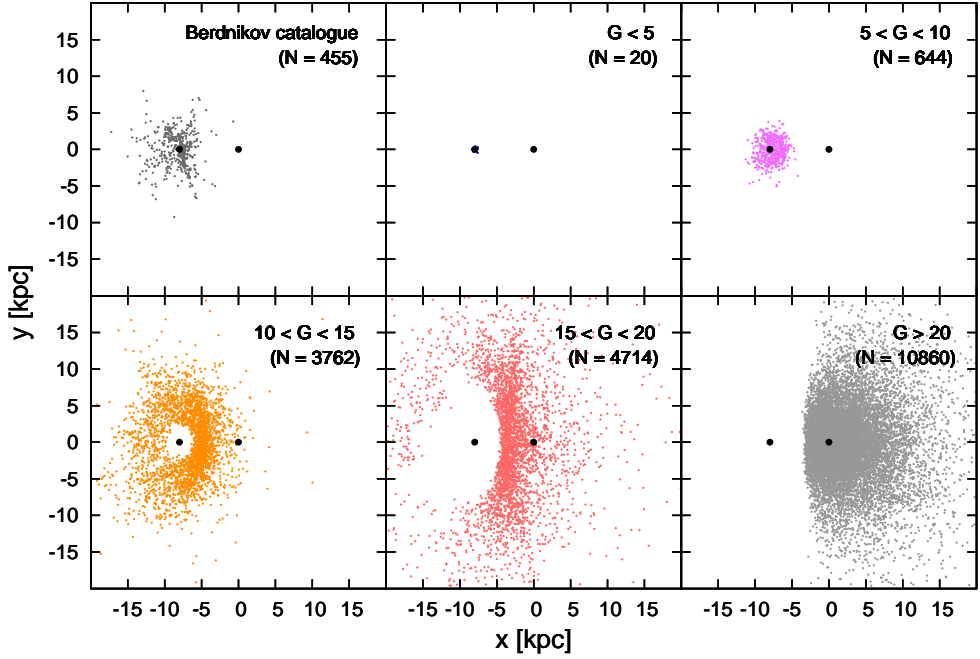}
  \caption{The Galactic Cepheids as observed from the Sun, with the Galaxy seen face on. The Sun is positioned at $(x,y) = (-8, 0)$ kpc, and the Galactic center at (0,0). The upper left panel shows the Berdnikov sample, and the other five panels show the synthetic Cepheid population divided into different $G$ magnitude bins. The Cepheids in the lower right panel and those brighter than $G=6$ would not be observed by Gaia.}
  \label{fig:galaxy}
\end{figure*}

\subsection{Gaia observations}

To simulate how the Cepheids are observed by Gaia we need their positions and apparent $G$ magnitudes as seen from the Sun. The positions are immediately obtained from the simulated galactocentric $(x,y,z)$ coordinates by subtracting the coordinates of the Sun. The apparent $V$ magnitude is given by 
\begin{equation}
	\label{eq:distmod}
	V = M_V + 5 \log r - 5 - A_V \, ,
\end{equation}
where $r$ is the heliocentric distance in pc and $A_V$ the total extinction. \citet{2005AJ....130..659A} describe an axisymmetric three-dimensional Galactic extinction model obtained from observations of the distribution of Galactic HI and H$_2$ gas. We use this model to obtain both $A_V$ and the colour excess in $V-I$, assuming the ratio
$R_{V-I}\equiv A_V/E_{V-I}=2.42$ \citep{2000asqu.book.....C,2003A&A...404..423T}. 
The Gaia $G$ band is similar to $V$ for blue objects but brighter for red objects. The $V$ magnitude is transformed into $G$ by the following relation from \citet{2010IAUS..261..296L}:
\begin{equation}
	G = V - 0.017 - 0.088(V-I) - 0.163(V-I)^2 + 0.009(V-I)^3
	\label{eq:G}
\end{equation} 
\citep[cf.][]{2010A&A...523A..48J}.
The synthetic population is then observed according to a standard model developed by the Gaia Data Processing and Analysis Consortium 
\citep[DPAC;][]{2008IAUS..248..224M}. Gaia has a bright limit of $G = 6$, where the detectors saturate, and a faint limit of $G = 20$, corresponding to $V \simeq 20$--25. The Gaia observational model predicts the standard error in the measured parallax, $\sigma_\varpi$, for each object in this range, taking into account the standard error per scan across the object, depending on the $G$ magnitude, and the number and geometry of the scans over the five year mission, depending on the object's position on the sky. The observed parallax ($\varpi_{\rm G}$) is then obtained by adding a normally distributed random measurement error, with the calculated $\sigma_\varpi$, to the true parallax.

\begin{figure}
	\centering
	\resizebox{0.9\hsize}{!}{\includegraphics[angle=270]{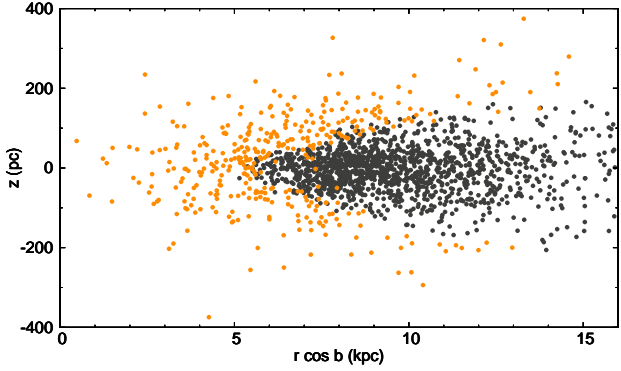}}
  \caption{The simulated inner Galaxy ($|l| < 5^{\circ}$), with projected distance to the Sun plotted versus height above the Galactic plane. The light dots correspond to Cepheids that will be observable by Gaia ($G < 20$) and the dark dots correspond to Cepheids that are too faint to be observed by Gaia ($G > 20$).}
  \label{fig:innergalaxy}
\end{figure}

In Fig.~\ref{fig:galaxy} the Cepheids of the Berdnikov catalogue are compared with the synthetic Cepheid population, divided into five bins in apparent $G$ magnitude. We estimate that Gaia will observe roughly $9\,000$ Galactic Cepheids, or almost half of the total population. This value is found to be relatively insensitive to variations in the Cepheid distribution parameters. We note that even though the extinction towards the Galactic Centre is believed to be very large, several hundred Cepheids are still visible near and even behind it. These stars are all found at $\left| z \right|$ of hundreds of pc, where the total extinction is  relatively small compared to in the plane. This can also be seen in Fig.~\ref{fig:innergalaxy}, where the vertical distribution of Cepheids towards the Galactic Centre ($|l| < 5^{\circ}$) is plotted. At distances larger than 5 kpc, there are no Cepheids visible in the Galactic plane.

\section{Analysis of simulated observations}
\label{sec:analysis}

With the models described in the previous section it is possible to generate a list of observed Galactic Cepheids together with the simulated data 
(e.g., $P$, $G$, $V-I$, $\varpi_{\rm G}$, $\sigma_\varpi$). In this section we describe the tools used to analyse the simulated data and the results of the parameter fitting.

\subsection{Parameter fitting}
\label{sec:fitting}

The idea is to use the observed parallaxes of nearby Cepheids to determine the P-L relation, and to use this P-L relation for more distant Cepheids to determine the parallax bias. 
To avoid circularity, we make a simultaneous fit of the P-L relation and the parallax zero point to all the data. Following \citet{2003A&A...403..993K} the fitting is made in parallax space, where the error distribution is symmetric around the true values. 
Statistically, this is equivalent to the method of reduced parallaxes 
\citep{2002MNRAS.337.1035F}, and allows to handle correctly that some observed parallaxes are negative due to measurement errors. 
Such negative observed parallaxes are statistically valid measurements, but cannot be converted to distances or absolute magnitudes. 
During the parameter fitting, all measured objects are therefore usable, avoiding a possible bias due to selection \citep{1997MNRAS.286L...1F}. 
Rewriting Eq.~(\ref{eq:distmod}) and inserting the P-L relation, we get the observation equation:
\begin{equation}
	\varpi_{\rm G}~[\mu\mbox{as}] = 10^{5 + 0.2[a\log P + b-V+A_V]}+c+\mbox{noise}\, ,
		\label{eq:paramfit}
\end{equation}
where $a$ and $b$ are the slope and zero point of the P-L relation, respectively. $c$ represents the global parallax zero point error, and is expected to be zero if Gaia works well. We can safely assume that Gaia will be able to measure $P$ and $V$ with negligible uncertainty. If we assume negligible intrinsic dispersion of the P-L relation and that $A_V$ is known exactly, we have the best-case scenario for parallax zero-point verification using Cepheids. Each data point is then weighted entirely depending on its formal parallax uncertainty, $\sigma_\varpi$. 

As a more realistic alternative, we introduce some uncertainty in the knowledge of the extinction value $A_V$ by assuming a constant uncertainty $\sigma_{A_V} = 0.05$~mag for all objects. This is pessimistic for the bright and nearby, low-extinction Cepheids, but probably optimistic for high-extinction Cepheids. We then generate assumed extinction values that are normally distributed around the true values, and take the total uncertainty in Eq.~(\ref{eq:paramfit}) to be
$\sigma'_\varpi = [ \sigma_{\varpi}^2 + (\varpi \sigma_{A_V} / 2.17)^2 ]^{0.5}$. This will lessen the importance of the nearby Cepheids, but might also introduce additional bias effects since the calculated uncertainty uses the observed parallax and not the true one.
As seen from Eq.~(\ref{eq:paramfit}), an intrinsic dispersion of the absolute 
magnitude in the P-L relation will have the same effect as random errors in the assumed 
$A_V$. We can therefore use these experiments also to conclude on how such a dispersion would affect the results.

To avoid the uncertainties associated with determining the extinction for each Cepheid, we also investigate the use of a reddening-free method equivalent to the use of the 
Wesenheit function $W=V-R_{V-I}(V-I)$ \citep{1991PASP..103..933M,2009ApJ...696.1498M},
where $R_{V-I}$ is the ratio of the total to selective extinction.
If the period--colour relation is $V-I = d \log P + e$, the observation equation then becomes
\begin{equation}
	\varpi_G~[\mu\mbox{as}] = 10^{5 + 0.2[ k_1 \log P + k_2 -V+ R_{V-I}(V-I) ]} + c
	+\mbox{noise}\, ,
	\label{eq:paramfit1}
\end{equation}
where $k_1 = a - d R_{V-I}$ and $k_2 = b - e R_{V-I}$. It is necessary to introduce $k_1$ and $k_2$ as the new unknowns since it is not possible to solve simultaneously for all four parameters $a$, $b$, $d$ and $e$. This method requires however that $R_{V-I}$ is known, and we investigate the effect of assuming a value of $R_{V-I}$ that is too large by 5\%.

Least-squares fitting using the Newton-Raphson iterative method gives the parameters $a$, $b$ (or $k_1$ and $k_2$) and $c$ along with their formal uncertainties arising from the known uncertainties in the measured parallaxes. Biases and the total uncertainties resulting from all modelled effects can be obtained after multiple realisations of the Cepheid data and parameter fitting.

\begin{figure}
\resizebox{\hsize}{!}{\includegraphics[angle=270]{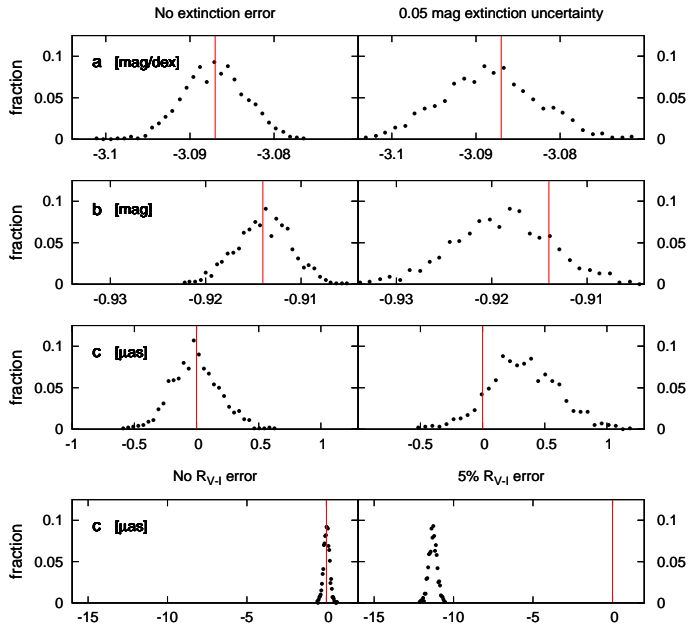}}
\caption{Distribution of the fitted parameters in 1000 realisations of the Cepheid data. The upper three panels show the distributions of $a$, $b$ (in the P-L relation) and $c$ (the parallax zero-point) for the model in Eq.~(\ref{eq:paramfit}) when extinction is perfectly known (left), and when it has an uncertainty of 0.05~mag (right). The bottom panels show the distribution of $c$ for the model in Eq.~(\ref{eq:paramfit1}) when $R_{V-I}$ is perfectly known (left) and when it has an error of 5\% (right). The vertical lines indicate the true values.}
\label{fig:paramdists}
\end{figure}

\subsection{Numerical results}\label{sec:numres}

In Fig.~\ref{fig:paramdists} we present the distributions of the fitted parameters $a$, $b$ and $c$ after $1000$ realisations of the Cepheid observational data. In the left panels, the correct extinction is assumed during the parameter fitting, meaning that the spread arises only due to the uncertainty in the Gaia parallaxes. This (unrealistic) best-case scenario  leads to very well-determined P-L relation parameters 
($\sigma_a=0.0035$~mag~dex$^{-1}$, $\sigma_b=0.0027$~mag) 
and a parallax zero-point uncertainty or $\sigma_c = 0.20~\mu$as. No bias is observed: the distributions are symmetric around the true parameter values. The bottom left panel shows the distribution of $c$ when using the reddening-free model of Eq.~(\ref{eq:paramfit1}) with the correct value of $R_{V-I}$. Again, the results are unbiased but the parallax zero-point uncertainty is slightly larger, $\sigma_c=0.25~\mu$as.

In the right panels of Fig.~\ref{fig:paramdists} we show how the results are affected by an imperfect knowledge of the extinction, all other factors being the same as in the left panels. The top three diagrams show the results for $a$, $b$ and $c$ when the assumed extinction has an uncertainty of $\sigma_{A_V}=0.05$~mag. The parallax zero-point uncertainty has increased to $\sigma_c = 0.27~\mu$as, and in addition there is a significant bias in the P-L relation zero-point ($b$) and a corresponding bias in the parallax zero-point of $0.3~\mu$as. The bottom right diagram shows that the reddening-free model of Eq.~(\ref{eq:paramfit1}) is very sensitive to an error in the assumed $R_{V-I}$. A 5\% error in $R_{V-I}$ introduces a bias of about $11~\mu$as in $c$. 
To keep the bias below $0.1~\mu$as would require that $R_{V-I}$ is known to better than 0.04\%. 

The intrinsic dispersion of the P-L relation is at least about 0.1~mag
(Sect.~\ref{sec:properties}), and its expected effect on $c$ is therefore at least twice
as big as the 0.05~mag uncertainty in the extinction, including a likely bias of the
order of $0.6~\mu$as. Using the reddening-free method reduces the scatter both
due to the extinction and the intrinsic dispersion, but instead the results are then 
very sensitive to an error in $R_{V-I}$, as we have seen.  

We have briefly investigated the effects of a metallicity dependent P-L relation. \citet{2008AA...488...25G} found $M_V = -2.60 \log P -1.30 + 0.27 [{\rm Fe/H}]$ for the Galaxy, where the metallicity dependence has an uncertainty of $0.30$~mag~dex$^{-1}$. 
We assumed this metallicity dependence to be the true one, and implemented a radial metallicity gradient of $[{\rm Fe/H}](R) = 0.42 - 0.052R$ with an internal scatter of 
$0.1$~dex \citep{2008A&A...490..613L}. Assuming that the metallicity of each Cepheid can be determined with an accuracy of $\sigma_{[{\rm Fe/H}]} = 0.1$~dex and adding a metallicity parameter in the P-L relation, the parallax zero-point uncertainty increases to $\sigma_c = 0.52~\mu$as in the case of a perfect knowledge of the extinction. This model also results in an increased bias of $0.45~\mu$as in $c$. If metallicity is not accounted for in the P-L model, the scatter in $c$ is somewhat reduced but the bias is even larger ($1.3~\mu$as).

The experiments described above used all the Cepheids observed by Gaia in the least-squares fitting, independent of their individual accuracies and degrees of extinction. 
As we have seen, the results are very sensitive to extinction errors. It is possible that this sensitivity is a consequence of including many Cepheids with large extinction in the analysis. In order to investigate this we tried various ways of removing the worst data, e.g., by only using Cepheids within certain distance or extinction limits. This gave only a slight improvement in terms of the scatter in $c$, but was usually found to introduce additional biases that proved difficult to avoid. One reason for this could be that the selection of `good' data depend on measured quantities, which in the case of noisy data invariably introduces selection biases. We also note that the method requires the observation of both nearby and distant Cepheids in order to separate the P-L zero point ($b$) from the parallax zero point ($c$); excluding either nearby or distant Cepheids from the analysed sample introduces large statistical uncertainties in both parameters.

\section{Conclusions}
\label{sec:conclusions}

In order to explore the full statistical potential of the Gaia parallaxes it is desirable that the global parallax zero point can be verified to within $0.1~\mu$as. We have explored the possible use of Galactic classical Cepheids for this purpose.

A model of the Galactic Cepheid population has been formulated which allows us to simulate their observation by the Gaia satellite. From the simulated data, we have made simultaneous fits of the P-L relation and the Gaia parallax zero point under a variety of assumptions.

We find that the parameters $a$ and $b$ of the P-L relation can be determined with a typical precision better than 0.01~mag~dex$^{-1}$ or 0.01~mag, respectively, which is far better than current calibrations. The results for the parallax zero point $c$ are less encouraging. Even under optimal circumstances (accurate knowledge of extinction, metallicity, etc), the Galactic Cepheid method cannot determine $c$ better than to within a few tenths of a $\mu$as. Moreover, we find that the resulting $c$ is very sensitive to errors in the extinction correction, or to an error in the $R_{V-I}$ value if a reddening-free method is used. Attempts to improve the situation, e.g., by limiting the sample to low-extinction Cepheids, were largely unsuccessful due to the introduction of additional biases caused by the selection being made from observed values.

By extrapolating Cepheid statistics from the \citet{2000A&AS..143..211B} catalogue, we 
estimate the total number of Galactic Cepheids to be $\sim\,$20\,000. We estimate that Gaia will observe about 9\,000 of them, which is a factor ten larger than the currently known number. Although many of them are faint, their observation by Gaia will greatly improve our knowledge of the P-L relation and its dependence of other factors such as metallicity. A detailed global modelling of their characteristics is very worthwhile, and should take into consideration a possible parallax zero point error. However, the ultimate astrophysical verification of Gaia's parallax zero-point is likely to depend on a combination of many different methods including the presented one.

Finally, if we assume that the parallax zero point can be verified to a good accuracy without the use of Cepheids, one could use the Cepheids observed by Gaia to learn more about extinction. With the highly accurate observations that Gaia will provide, this method could yield very precise mapping of the Galactic extinction ($R$ and $A_{\rm G}$) with an accuracy that has previously not been achievable.

\begin{acknowledgements}
We like to thank Berry Holl, Anthony Brown, Timo Prusti and the referee, Floor van Leeuwen, for helpful comments.
\end{acknowledgements}

\bibliography{refs}

\begin{thebibliography}{43}
\expandafter\ifx\csname natexlab\endcsname\relax\def\natexlab#1{#1}\fi

\bibitem[{{Am{\^o}res} \& {L{\'e}pine}(2005)}]{2005AJ....130..659A}
{Am{\^o}res}, E.~B. \& {L{\'e}pine}, J.~R.~D. 2005, \aj, 130, 659

\bibitem[{{Berdnikov} {et~al.}(2000){Berdnikov}, {Dambis}, \&
  {Vozyakova}}]{2000A&AS..143..211B}
{Berdnikov}, L.~N., {Dambis}, A.~K., \& {Vozyakova}, O.~V. 2000, \aaps, 143,
  211

\bibitem[{{Binney} \& {Merrifield}(1998)}]{1998gaas.book.....B}
{Binney}, J. \& {Merrifield}, M. 1998, {Galactic astronomy} (Princeton
  University Press)

\bibitem[{{Cox}(2000)}]{2000asqu.book.....C}
{Cox}, A.~N. 2000, {Allen's astrophysical quantities}, ed. {Cox, A.~N.}

\bibitem[{{Feast}(2002)}]{2002MNRAS.337.1035F}
{Feast}, M. 2002, \mnras, 337, 1035

\bibitem[{{Feast} \& {Catchpole}(1997)}]{1997MNRAS.286L...1F}
{Feast}, M.~W. \& {Catchpole}, R.~M. 1997, \mnras, 286, L1

\bibitem[{{Fouqu{\'e}} {et~al.}(2007){Fouqu{\'e}}, {Arriagada}, {Storm},
  {Barnes}, {Nardetto}, {M{\'e}rand}, {Kervella}, {Gieren}, {Bersier},
  {Benedict}, \& {McArthur}}]{2007AA...476...73F}
{Fouqu{\'e}}, P., {Arriagada}, P., {Storm}, J., {et~al.} 2007, \aap, 476, 73

\bibitem[{{Freedman} {et~al.}(2001){Freedman}, {Madore}, {Gibson}, {Ferrarese},
  {Kelson}, {Sakai}, {Mould}, {Kennicutt}, {Ford}, {Graham}, {Huchra},
  {Hughes}, {Illingworth}, {Macri}, \& {Stetson}}]{2001ApJ...553...47F}
{Freedman}, W.~L., {Madore}, B.~F., {Gibson}, B.~K., {et~al.} 2001, \apj, 553,
  47

\bibitem[{{Groenewegen}(2008)}]{2008AA...488...25G}
{Groenewegen}, M.~A.~T. 2008, \aap, 488, 25

\bibitem[{{Holl} {et~al.}(2010){Holl}, {Hobbs}, \&
  {Lindegren}}]{2010IAUS..261..320H}
{Holl}, B., {Hobbs}, D., \& {Lindegren}, L. 2010, in IAU Symposium, Vol. 261,
  IAU Symposium, ed. {S.~A.~Klioner, P.~K.~Seidelmann, \& M.~H.~Soffel},
  320--324

\bibitem[{{Jordi} {et~al.}(2010){Jordi}, {Gebran}, {Carrasco}, {de Bruijne},
  {Voss}, {Fabricius}, {Knude}, {Vallenari}, {Kohley}, \&
  {Mora}}]{2010A&A...523A..48J}
{Jordi}, C., {Gebran}, M., {Carrasco}, J.~M., {et~al.} 2010, \aap, 523, A48+

\bibitem[{{Knapp} {et~al.}(2003){Knapp}, {Pourbaix}, {Platais}, \&
  {Jorissen}}]{2003A&A...403..993K}
{Knapp}, G.~R., {Pourbaix}, D., {Platais}, I., \& {Jorissen}, A. 2003, \aap,
  403, 993

\bibitem[{{Lane} {et~al.}(2002){Lane}, {Creech-Eakman}, \&
  {Nordgren}}]{2002ApJ...573..330L}
{Lane}, B.~F., {Creech-Eakman}, M.~J., \& {Nordgren}, T.~E. 2002, \apj, 573,
  330

\bibitem[{{Leavitt}(1908)}]{1908AnHar..60...87L}
{Leavitt}, H.~S. 1908, Annals of Harvard College Observatory, 60, 87

\bibitem[{{Leavitt} \& {Pickering}(1912)}]{1912HarCi.173....1L}
{Leavitt}, H.~S. \& {Pickering}, E.~C. 1912, Harvard College Observatory
  Circular, 173, 1

\bibitem[{{Lemasle} {et~al.}(2008){Lemasle}, {Fran{\c c}ois}, {Piersimoni},
  {Pedicelli}, {Bono}, {Laney}, {Primas}, \&
  {Romaniello}}]{2008A&A...490..613L}
{Lemasle}, B., {Fran{\c c}ois}, P., {Piersimoni}, A., {et~al.} 2008, \aap, 490,
  613

\bibitem[{{Li} {et~al.}(2002){Li}, {Harrington}, \&
  {Borkowski}}]{2002AJ....123.2676L}
{Li}, J., {Harrington}, J.~P., \& {Borkowski}, K.~J. 2002, \aj, 123, 2676

\bibitem[{{Lindegren}(2005)}]{2005ESASP.576...29L}
{Lindegren}, L. 2005, in ESA Special Publication, Vol. 576, The
  Three-Dimensional Universe with Gaia, ed. {C.~Turon, K.~S.~O'Flaherty, \&
  M.~A.~C.~Perryman}, 29--34

\bibitem[{{Lindegren}(2010)}]{2010IAUS..261..296L}
{Lindegren}, L. 2010, in IAU Symposium, ed. {S.~A.~Klioner, P.~K.~Seidelmann,
  \& M.~H.~Soffel}, Vol. 261, 296--305

\bibitem[{{Lindegren} {et~al.}(2008){Lindegren}, {Babusiaux}, {Bailer-Jones},
  {Bastian}, {Brown}, {Cropper}, {H{\o}g}, {Jordi}, {Katz}, {van Leeuwen},
  {Luri}, {Mignard}, {de Bruijne}, \& {Prusti}}]{2008IAUS..248..217L}
{Lindegren}, L., {Babusiaux}, C., {Bailer-Jones}, C., {et~al.} 2008, in IAU
  Symposium, ed. {W.~J.~Jin, I.~Platais, \& M.~A.~C.~Perryman}, Vol. 248,
  217--223

\bibitem[{{Lindegren} \& {de Bruijne}(2005)}]{2005ASPC..338...25L}
{Lindegren}, L. \& {de Bruijne}, J.~H.~J. 2005, in Astronomical Society of the
  Pacific Conference Series, Vol. 338, Astrometry in the Age of the Next
  Generation of Large Telescopes, ed. {P.~K.~Seidelmann \& A.~K.~B.~Monet},
  25--36

\bibitem[{{Madore} \& {Freedman}(1991)}]{1991PASP..103..933M}
{Madore}, B.~F. \& {Freedman}, W.~L. 1991, \pasp, 103, 933

\bibitem[{{Madore} \& {Freedman}(2009)}]{2009ApJ...696.1498M}
{Madore}, B.~F. \& {Freedman}, W.~L. 2009, \apj, 696, 1498

\bibitem[{{Majaess} {et~al.}(2009){Majaess}, {Turner}, \&
  {Lane}}]{2009MNRAS.398..263M}
{Majaess}, D.~J., {Turner}, D.~G., \& {Lane}, D.~J. 2009, \mnras, 398, 263

\bibitem[{{McKee} \& {Williams}(1997)}]{1997ApJ...476..144M}
{McKee}, C.~F. \& {Williams}, J.~P. 1997, \apj, 476, 144

\bibitem[{{McLaughlin} {et~al.}(2006){McLaughlin}, {Anderson}, {Meylan},
  {Gebhardt}, {Pryor}, {Minniti}, \& {Phinney}}]{2006ApJS..166..249M}
{McLaughlin}, D.~E., {Anderson}, J., {Meylan}, G., {et~al.} 2006, \apjs, 166,
  249

\bibitem[{{Mignard}(2011)}]{2011AdSpR..47..356M}
{Mignard}, F. 2011, Advances in Space Research, 47, 356

\bibitem[{{Mignard} {et~al.}(2008){Mignard}, {Bailer-Jones}, {Bastian},
  {Drimmel}, {Eyer}, {Katz}, {van Leeuwen}, {Luri}, {O'Mullane}, {Passot},
  {Pourbaix}, \& {Prusti}}]{2008IAUS..248..224M}
{Mignard}, F., {Bailer-Jones}, C., {Bastian}, U., {et~al.} 2008, in IAU
  Symposium, Vol. 248, IAU Symposium, ed. {W.~J.~Jin, I.~Platais, \&
  M.~A.~C.~Perryman}, 224--230

\bibitem[{{Ngeow} {et~al.}(2009){Ngeow}, {Kanbur}, {Neilson}, {Nanthakumar}, \&
  {Buonaccorsi}}]{2009ApJ...693..691N}
{Ngeow}, C., {Kanbur}, S.~M., {Neilson}, H.~R., {Nanthakumar}, A., \&
  {Buonaccorsi}, J. 2009, \apj, 693, 691

\bibitem[{{Olling}(2007)}]{2007MNRAS.378.1385O}
{Olling}, R.~P. 2007, \mnras, 378, 1385

\bibitem[{{Perryman} {et~al.}(2001){Perryman}, {de Boer}, {Gilmore}, {H{\o}g},
  {Lattanzi}, {Lindegren}, {Luri}, {Mignard}, {Pace}, \& {de
  Zeeuw}}]{2001A&A...369..339P}
{Perryman}, M.~A.~C., {de Boer}, K.~S., {Gilmore}, G., {et~al.} 2001, \aap,
  369, 339

\bibitem[{{Romaniello} {et~al.}(2008){Romaniello}, {Primas}, {Mottini},
  {Pedicelli}, {Lemasle}, {Bono}, {Fran{\c c}ois}, {Groenewegen}, \&
  {Laney}}]{2008A&A...488..731R}
{Romaniello}, M., {Primas}, F., {Mottini}, M., {et~al.} 2008, \aap, 488, 731

\bibitem[{{Salpeter}(1955)}]{1955ApJ...121..161S}
{Salpeter}, E.~E. 1955, \apj, 121, 161

\bibitem[{{Sandage} {et~al.}(2004){Sandage}, {Tammann}, \&
  {Reindl}}]{2004AA...424...43S}
{Sandage}, A., {Tammann}, G.~A., \& {Reindl}, B. 2004, \aap, 424, 43

\bibitem[{{Schaefer}(2008)}]{2008AJ....135..112S}
{Schaefer}, B.~E. 2008, \aj, 135, 112

\bibitem[{{Tammann} {et~al.}(2003){Tammann}, {Sandage}, \&
  {Reindl}}]{2003A&A...404..423T}
{Tammann}, G.~A., {Sandage}, A., \& {Reindl}, B. 2003, \aap, 404, 423

\bibitem[{{Torres} {et~al.}(1997){Torres}, {Stefanik}, \&
  {Latham}}]{1997ApJ...485..167T}
{Torres}, G., {Stefanik}, R.~P., \& {Latham}, D.~W. 1997, \apj, 485, 167

\bibitem[{{Trimble}(1973)}]{1973PASP...85..579T}
{Trimble}, V. 1973, \pasp, 85, 579

\bibitem[{{Udalski} {et~al.}(1999){Udalski}, {Szymanski}, {Kubiak},
  {Pietrzynski}, {Soszynski}, {Wozniak}, \& {Zebrun}}]{1999AcA....49..201U}
{Udalski}, A., {Szymanski}, M., {Kubiak}, M., {et~al.} 1999, \actaa, 49, 201

\bibitem[{{van de Ven} {et~al.}(2006){van de Ven}, {van den Bosch}, {Verolme},
  \& {de Zeeuw}}]{2006A&A...445..513V}
{van de Ven}, G., {van den Bosch}, R.~C.~E., {Verolme}, E.~K., \& {de Zeeuw},
  P.~T. 2006, \aap, 445, 513

\bibitem[{{van der Kruit}(1988)}]{1988A&A...192..117V}
{van der Kruit}, P.~C. 1988, \aap, 192, 117

\bibitem[{{van Leeuwen}(2005)}]{2005A&A...439..805V}
{van Leeuwen}, F. 2005, \aap, 439, 805

\bibitem[{{Williams} \& {McKee}(1997)}]{1997ApJ...476..166W}
{Williams}, J.~P. \& {McKee}, C.~F. 1997, \apj, 476, 166

\end{thebibliography}
\bibliographystyle{aa}

\end{document}